\documentclass[letters, fleqn, usenatbib]{mnras}
\usepackage{amssymb,amsmath}
\usepackage{graphicx}
\usepackage{xcolor}
\usepackage{multirow}
\usepackage[T1]{fontenc}
\usepackage{ae,aecompl}
\usepackage{newtxtext,newtxmath}
\usepackage{longtable,listings}

\def\oii{[O~{\sc ii}]$\lambda3727$\AA\ }
\def\msig{$M_{\rm BH}-\sigma$\ }

\def\obj{SDSS J1619}

\title[Spectral decomposition of SDSS J1619]{SDSS J1619 with blue-shifted broad components in H$\alpha$ and in [O~{\sc iii}] 
having similar line width and velocity shifts: a recoiling SMBH candidate?}

\author[Zhang]
{Xue-Guang Zhang\thanks{Corresponding author Email: \href{mailto:xgzhang@gxu.edu.cn}{xgzhang@gxu.edu.cn}}\\
Guangxi Key Laboratory for Relativistic Astrophysics, School of Physical Science and Technology, GuangXi University, 
Nanning, 530004, P. R. China}

\pubyear{2023}

\begin{document}

\label{firstpage}

\pagerange{\pageref{firstpage}--\pageref{lastpage}}

\maketitle


\begin{abstract} 
	In this Letter, we report a potential candidate of recoiling supermassive black hole (rSMBH) in SDSS J1619 based on 
similar velocity shifts and line widths of the blue-shifted broad components in H$\alpha$ and [O~{\sc iii}] doublet. 
The measured line width ratio between blue-shifted broad H$\alpha$ and broad [O~{\sc iii}] line is 1.06, if compared with common 
values around 5.12 for normal Type-1 AGN, indicating different properties of the blue-shifted broad components in SDSS J1619 
from those of normal QSOs. The virial BH mass $M_{BHr}$ derived from the broad H$\alpha$ is consistent with the mass expected 
from the \msig~ relation. The similar velocity shifts and line widths of the blue-shifted broad components in H$\alpha$ and 
[O~{\sc iii}] and the virial BH mass derived from the H$\alpha$ broad line emissions that is consistent with the mass expected 
from the \msig~ relation, can be explained by a rSMBH scenario. Besides the rSMBH scenario, either the similar line widths of 
the blue-shifted broad components in H$\alpha$ and in [O~{\sc iii}] or the consistency between the virial BH mass and the mass 
expected from the \msig~ relation cannot be explained by the other proposed models in SDSS J1619.
\end{abstract}


\section{Introduction}

	A black hole (BH) can be kicked away from central region of an active galactic nuclei (AGN), due to gravitational wave 
carrying away linear momentum, as discussed in \citet{ms06, vm07, bl08, km08b, gm08, bs16, zy17, sh19}. The BH being kicked 
away is also called as gravitational wave recoiling supermassive BH (=rSMBH), leading to the known off-nucleus AGN with shifted 
broad emission lines due to materials in broad emission line regions (BLRs) bound to the recoiling BH. Until now, tens of AGN 
have been reported with blue-shifted broad lines.

	SDSS J0927+2943 at redshift 0.713 has been firstly reported in \citet{kz08} with rSMBH expected blue-shifted velocity 
2650km/s in low/high-ionization broad emission lines. However, a binary black hole (BBH) system can also lead to the blue-shifted 
lines in SDSS J0927+2943, as discussed in \citet{be09}. SDSS J1050 has been reported in \citet{sr09} with blue-shifted velocity 
3500km/s in broad H$\beta$, however, BLRs lying into central accretion disk (disk emitter) would be preferred in SDSS J1050 to 
explain the shifted broad H$\beta$. Similar as the results in SDSS J0927+2943 and in SDSS J1050, there are blue-shifted broad 
lines reported in several individual AGN, such as SDSS J0956+5128\ in \citet{ss12}, CXO J1015+6259\ in \citet{ky17}, SDSS 
J1056+5516\ in \citet{ks17}, Mrk1018 in \citet{ky18}, 3C186 in \citet{ce17, ce18}, J0437+2456\ in \citet{ps21}, etc., but the 
rSMBH scenario cannot be accepted as the unique scenario to explain the reported blue-shifted emission lines. 
Meanwhile, besides reported individual AGN with blue-shifted broad emission lines, there are about 88 low redshift ($z<0.7$) 
SDSS quasars reported with blue-shifted velocities larger than 1000km/s in broad H$\beta$ in \citet{eb12, re15, re17}, and 
rather than the rSMBH scenario, BBH systems would be preferred in fraction of the low redshift quasars. Therefore, not only 
rSMBH scenario, but also BBH system, disk emitter or probable outflowing model can be applied to explain shifted broad emission 
lines in AGN.

	Currently, there are probably two main reasons which affect the plausibility of the rSMBH scenario. First, when a rSMBH 
is kicked away in AGN, probably not total materials in central BLRs are carried away with the rSMBH, but part of BLRs should 
be probably left in central region of AGN. In other words, there are two components in observed broad Balmer emission lines, 
one component from the BLRs bound to the rSMBH and the other component related to the materials in BLRs left in central region, 
leading to more complicated profiles of observed broad emission lines which not only include contributions from rSMBH expected 
blue-shifted broad components but also include rSMBH-independent components. How to effectively ignore effects of rSMBH-independent 
components is a challenge on studying properties of rSMBH expected blue-shifted broad emission features. However, considering 
the rSMBH-independent broad emission components coming from the left part of BLRs in central regions of AGN, it will be preferred 
to detect and study rSMBH-expected blue-shifted broad emission lines in Type-1.9 AGN, because rSMBH-independent broad emission 
components can be heavily obscured as many as possible by central dust torus. Second, as discussed in \citet{ms06, km08b, gm08, 
bl11, bs16}, the rSMBH may be off-nucleus for $10^{6-9}$ years with 1pc-1kpc distance from the center. Meanwhile, as discussed 
in \citet{liu13, ha13, fk18, dz18, zh22a}, narrow emission line regions (NLRs) of AGN extend up to 1 kpc (NLRs sizes) from the 
central BHs. Therefore, a rSMBH with bounded materials related to central BLRs can reach NLRs in AGN, leading part of emission 
materials in NLRs to co-move with the rSMBH, strongly indicating that narrow emission lines should have similar shifted velocities 
and similar line widths as those of rSMBH expected blue-shifted broad emission lines. Unfortunately, there are so far no reports 
on effects of rSMBH on properties of narrow emission lines from NLRs. Considering the two main reasons above, it is interesting 
to detect blue-shifted components both in BLRs and NLRs.

	Therefore, in this Letter, a special Type-1.9 AGN, SDSS J161950.67+500535.31 (=SDSS J1619), is reported with similar line 
widths and similar velocity shifts of blue-shifted broad components in both [O~{\sc iii}] doublet and H$\alpha$, suggesting a 
rSMBH scenario in SDSS J1619. This Letter is organized as follows. Section 2 and 3 show the analysis of the spectrum and the 
necessary discussions. Section 4 shows our final conclusions. And in this Letter, the cosmological parameters of 
$H_{0}~=~70{\rm km\cdot s}^{-1}{\rm Mpc}^{-1}$, $\Omega_{\Lambda}~=~0.7$ and $\Omega_{m}~=~0.3$ have been adopted.

\begin{figure*}
\centering\includegraphics[width = 18cm,height=5cm]{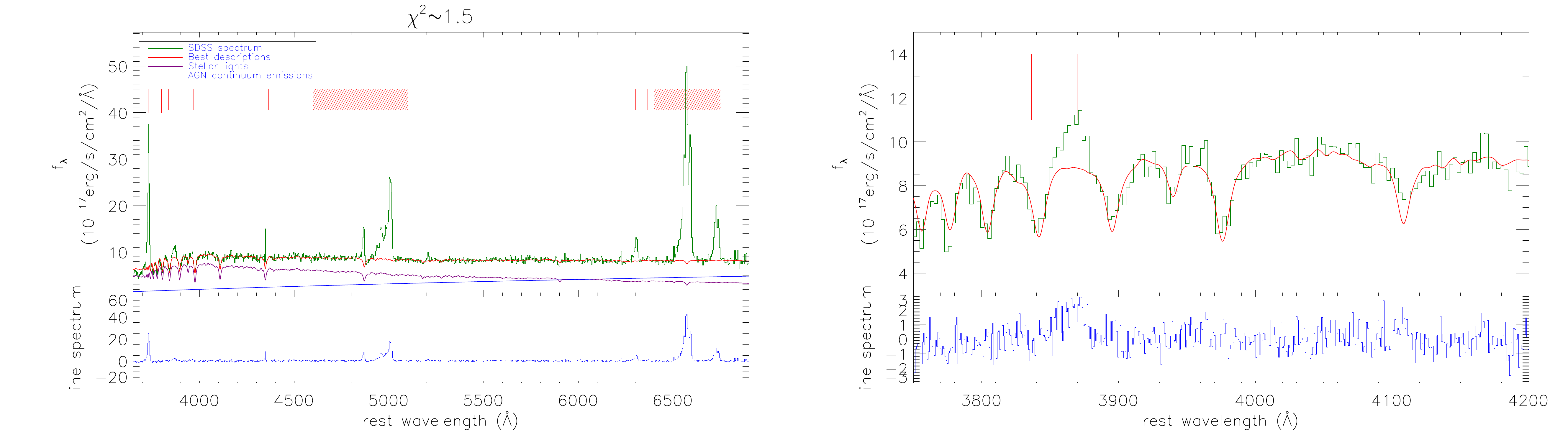}
\caption{Left panel shows the SDSS spectrum (in dark green) of \obj~ and the SSP method determined best descriptions (solid red 
line) to the spectrum with emission lines being masked out. In top region, as shown legend in top-left corner, solid purple 
line shows the SSP method determined stellar lights, solid blue line shows the continuum emissions, vertical red lines from left 
to right mark the following emission features masked out, including \oii, H$\theta$, H$\eta$, [Ne~{\sc iii}]$\lambda3869$\AA, 
He~{\sc i}$\lambda3891$\AA, Calcium K line, [Ne~{\sc iii}]$\lambda3968$\AA, Calcium H line, [S~{\sc ii}]$\lambda4070$\AA, H$\delta$, 
H$\gamma$, [O~{\sc iii}]$\lambda4364$\AA, He~{\sc i}$\lambda5877$\AA\ and [O~{\sc i}]$\lambda6300,6363$\AA, respectively, and the 
area filled by red lines around 5000\AA\ shows the region masked out including the emission features of probable He~{\sc ii}, 
broad and narrow H$\beta$ and [O~{\sc iii}] doublet, and the area filled by red lines around 6550\AA\ shows the region masked out 
including the emission features of broad and narrow H$\alpha$, [N~{\sc ii}] and [S~{\sc ii}] doublets. Bottom region of left panel 
shows the line spectrum calculated by the SDSS spectrum minus sum of the stellar lights and the continuum emissions. Right panels 
show the best descriptions to the absorption features around 4000\AA~ (rest wavelength range from 3750 to 4200\AA) (top right panel) 
and the corresponding line spectrum (the SDSS spectrum minus sum of the stellar lights and the continuum emissions) (bottom right 
panel). In right panels, line styles and symbols have the same meanings as those in the left panels. In top right panel, vertical 
red lines from left to right mark the following emission features of H$\theta$, H$\eta$, [Ne~{\sc iii}]$\lambda3869$\AA, 
He~{\sc i}$\lambda3891$\AA, Calcium K line, [Ne~{\sc iii}]$\lambda3968$\AA, Calcium H line, [S~{\sc ii}]$\lambda4070$\AA, H$\delta$.}
\label{spec}
\end{figure*}

\section{Spectroscopic results of SDSS J1619}

	Motivated by our previous measurements of opening angle of central dust tours in a Type-1.9 AGN with double-peaked broad 
H$\alpha$ in \citet{zh23}, on studying properties of Type-1.9 AGN is one of our ongoing projects. Among our sample of Type-1.9 AGN, 
\obj~ is selected as the subject of this Letter, due to its apparent blue-shifted broad components in both Balmer emission lines 
and forbidden [O~{\sc iii}] lines. 

\begin{table}
\caption{Line parameters of each Gaussian emission component}
\begin{tabular}{llll}
\hline\hline
line & $\lambda_0$ & $\sigma$ & flux \\
\hline\hline
Broad H$\alpha$  & 6549.52$\pm$5.61 & 19.76$\pm$3.35 & 284$\pm$68 \\
\hline
Broad H$\beta$  &  4851.51$\pm$4.16  & 14.64$\pm$2.48  & 23$\pm$8 \\
\hline\hline
\multirow{2}{*}{Narrow H$\alpha$} & 6571.31$\pm$0.12 & 4.85$\pm$0.16 &  537$\pm$45 \\
& 6559.15$\pm$0.28   & 2.29$\pm$0.31  & 51$\pm$10 \\
\hline
\multirow{2}{*}{Narrow H$\beta$} & 4867.64$\pm$0.09 & 3.59$\pm$0.12 &  81$\pm$4 \\
&  4858.63$\pm$0.21 & 1.70$\pm$0.23 & 7$\pm$2 \\
\hline
\multirow{3}{*}{[O~{\sc iii}]$\lambda5007$\AA} & 5012.63$\pm$0.15 & 3.89$\pm$0.31 & 121$\pm$15 \\
& 5004.44$\pm$0.37 & 3.25$\pm$0.24 & 100$\pm$13 \\
& 4991.86$\pm$0.79 & 14.22$\pm$0.56 & 272$\pm$12 \\
\hline
\multirow{2}{*}{[N~{\sc ii}]$\lambda6583$\AA} & 6591.80$\pm$0.46 & 4.55$\pm$0.62 & 223$\pm$100 \\
& 6587.23$\pm$3.28 & 8.18$\pm$1.65 & 208$\pm$96 \\
\hline\hline
\end{tabular}\\
{\bf Notice:} The first column shows which line is measured. The Second, third, fourth columns show the measured line parameters: 
the center wavelength $\lambda_0$ in units of \AA, the line width (second moment) $\sigma$ in units of \AA~ and the line flux in 
units of ${\rm 10^{-17}~erg/s/cm^2}$.
\end{table}

	SDSS spectrum (plate-mjd-fiberid=2884-54526-0145) of \obj~ at $z\sim0.283$ is collected from SDSS DR16 \citep{ap21} and 
shown in Fig.~\ref{spec} with median signal-to-noise about 12. The redshift is determined by the SDSS 
pipeline\footnote{\url{https://www.sdss3.org/dr8/algorithms/redshifts.php}}. Due to apparent stellar absorption features, host galaxy 
contributions should be firstly determined and subtracted, in order to measure properties of narrow/broad emission lines. Based 
on the 39 simple stellar population templates from \citet{bc03, ka03}, the SSP (simple Stellar Population) method discussed 
and accepted in \citet{cm05, cm17, zh21a, zh21b, zh22a, zh22b} is applied to determine the host galaxy contributions. Meanwhile, 
besides the stellar templates, a fourth order polynomial function is applied to describe intrinsic AGN continuum emissions. Left 
panels of Fig.~\ref{spec} show the best descriptions to the SDSS spectrum (with emission lines being masked out) of \obj~ and the 
corresponding line spectrum (the SDSS spectrum minus the best descriptions) through the Levenberg-Marquardt least-squares 
minimization technique (the known MPFIT package), with $\chi^2/dof\sim1.5$ (summed squared residuals divided by degree of freedom). 
Right panels of Fig.~\ref{spec} shows clear descriptions to the absorption features around 4000\AA. The determined stellar velocity 
dispersion (the broadening velocity for the stellar templates) is about 165$\pm$29{\rm km/s}. Meanwhile, accepted the SDSS pipeline 
determined redshift 0.283, an additional shifted velocity about 500$\pm$20${\rm km/s}$ for the stellar templates is needed to 
describe the stellar absorption features.

\begin{figure*}
\centering\includegraphics[width = 18cm,height=5cm]{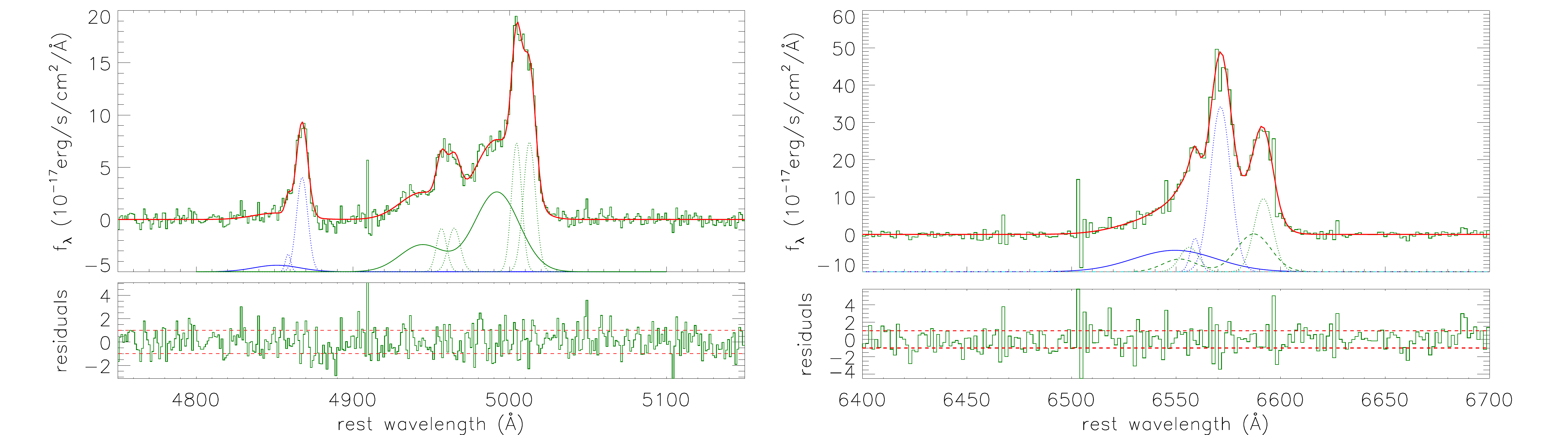}
\caption{Top panels show the best fitting results (solid red line) to the emission lines in the line spectrum (solid dark green 
line), bottom panels show the corresponding residuals calculated by the line spectrum minus the best fitting results 
and then divided by uncertainties of the SDSS spectrum. In top left panel, dotted blue lines show the two components in narrow 
H$\beta$, dotted line in dark green and solid line in dark green show the double-peaked features and the blue-shifted broad 
component related to wings in [O~{\sc iii}] doublet, solid blue line shows the broad component in H$\beta$. In top right panel, 
dotted blue lines show the two components in narrow H$\alpha$, solid blue line shows the blue-shifted broad component in H$\alpha$, 
dotted lines in dark green and dashed lines in dark green show the core components and the blue-shifted component related to wings 
in [N~{\sc ii}] doublet. In each bottom panel, horizontal dashed lines show residuals=$\pm1$, respectively.
}
\label{line}
\end{figure*}

	After subtractions of the stellar lights and the continuum emissions, emission lines around H$\beta$ (from 4750 to 
5150\AA~ in rest frame) and around H$\alpha$ (from 6400 to 6700\AA~ in rest frame) can be measured, similar as what we have 
recently done in \citet{zh21a, zh21b, zh22a, zh22b, zh22c}. Two broad Gaussian functions are applied to describe broad component 
in H$\beta$ (H$\alpha$). Two Gaussian functions plus one another Gaussian function are applied to describe the core components 
(with probable double-peaked features) and the component related to blue-shifted wing in each narrow emission line, including 
[O~{\sc iii}]$\lambda4959,5007$\AA~ doublet, narrow H$\beta$, [N~{\sc ii}]$\lambda6548,6583$\AA~ doublet and narrow H$\alpha$. 
When the model functions above are applied, the following restrictions are accepted. First, corresponding components of the 
[O~{\sc iii}] doublet (the [N~{\sc ii}] doublet, or the narrow Balmer lines) have the same redshift and the same line widths in 
velocity space. Second, corresponding components in the broad Balmer lines have the same redshift and the same line widths in 
velocity space. Third, flux ratio of the [O~{\sc iii}] doublet (the [N~{\sc ii}] doublet) is fixed to the theoretical value 3. 
Fourth, each Gaussian component has intensity not smaller than zero. Then, Fig.~\ref{line} shows the best descriptions to the 
emission lines of \obj, through the MPFIT package, with $\chi^2/dof\sim1.6$. The measured line parameters are listed in Table~1.

	Based on the best fitting results, two points can be found. First, the observed flux ratio (Balmer Decrement) of broad 
H$\alpha$ to broad H$\beta$ is about $12.35_{-5.38}^{+11.12}$ in \obj, quite larger than common values around 3.1 as shown in 
\citet{van01, dw08} in normal Type-1 AGN, indicating \obj~ can be classified as a Type-1.9 AGN. Second, blue-shifted 
broad components can be found in H$\alpha$ and also in forbidden [O~{\sc iii}] doublet. Moreover, due to double-peaked features 
in [O~{\sc iii}]$\lambda5007$\AA, it is hard to calculate velocity shifts of the broad component in [O~{\sc iii}]$\lambda5007$\AA~ 
relative to the narrow component in [O~{\sc iii}]$\lambda5007$\AA. Therefore, accepted the theoretical vacuum wavelength 6564.61\AA~ 
and 5008.24\AA~ for H$\alpha$ and [O~{\sc iii}]$\lambda5007$\AA, the blue-shifted broad component in H$\alpha$ has shifted velocity 
690$\pm$240km/s (calculated by central wavelength difference between the theoretical vacuum value and the measured value in Table~1) 
and second moment 900$\pm$170km/s, which are not quite different from the shifted velocity 980$\pm$50km/s and second moment 
850$\pm$34km/s of the blue-shifted broad component in [O~{\sc iii}] doublet, after considering uncertainties of the values. And 
the uncertainties of the velocity shifts and the second moments are calculated by the measured uncertainties (listed in Table~1) 
of the central wavelengths and the second moments of the broad blue-shifted broad component in H$\alpha$ and in 
[O~{\sc iii}]$\lambda5007$\AA.

	Meanwhile, due to single-peaked feature in [N~{\sc ii}]$\lambda6583$\AA, relative to the central wavelength of the 
core component in [N~{\sc ii}]$\lambda6583$\AA, the shifted broad extended component with second moment about 370$\pm$75km/s and 
velocity shift about 208$\pm$170km/s can be found in [N~{\sc ii}] doublet, quite smaller than the second moments and 
the velocity shifts of the blue-shifted broad components in H$\alpha$ and in [O~{\sc iii}], indicating the blue-shifted 
components in [N~{\sc ii}] doublet have quite different kinematic properties from those of the blue-shifted broad components in 
H$\alpha$ and in [O~{\sc iii}]. Certainly, if accepted the theoretical vacuum wavelength 6585.27\AA~ for 
[N~{\sc ii}]$\lambda6583$\AA, the extended component in [N~{\sc ii}] doublet should be red-shifted component, totally different 
properties from those of the blue-shifted broad components in H$\alpha$ and in [O~{\sc iii}]. Therefore, the shifted 
broad components in [N~{\sc ii}] doublet probably due to local kinematic properties of NLRs. There are no further discussions on 
the shifted components in [N~{\sc ii}], besides the fitting results shown in Fig~\ref{line} and the line parameters 
listed in Table~1. 

\begin{figure}
\centering\includegraphics[width = 8cm,height=5cm]{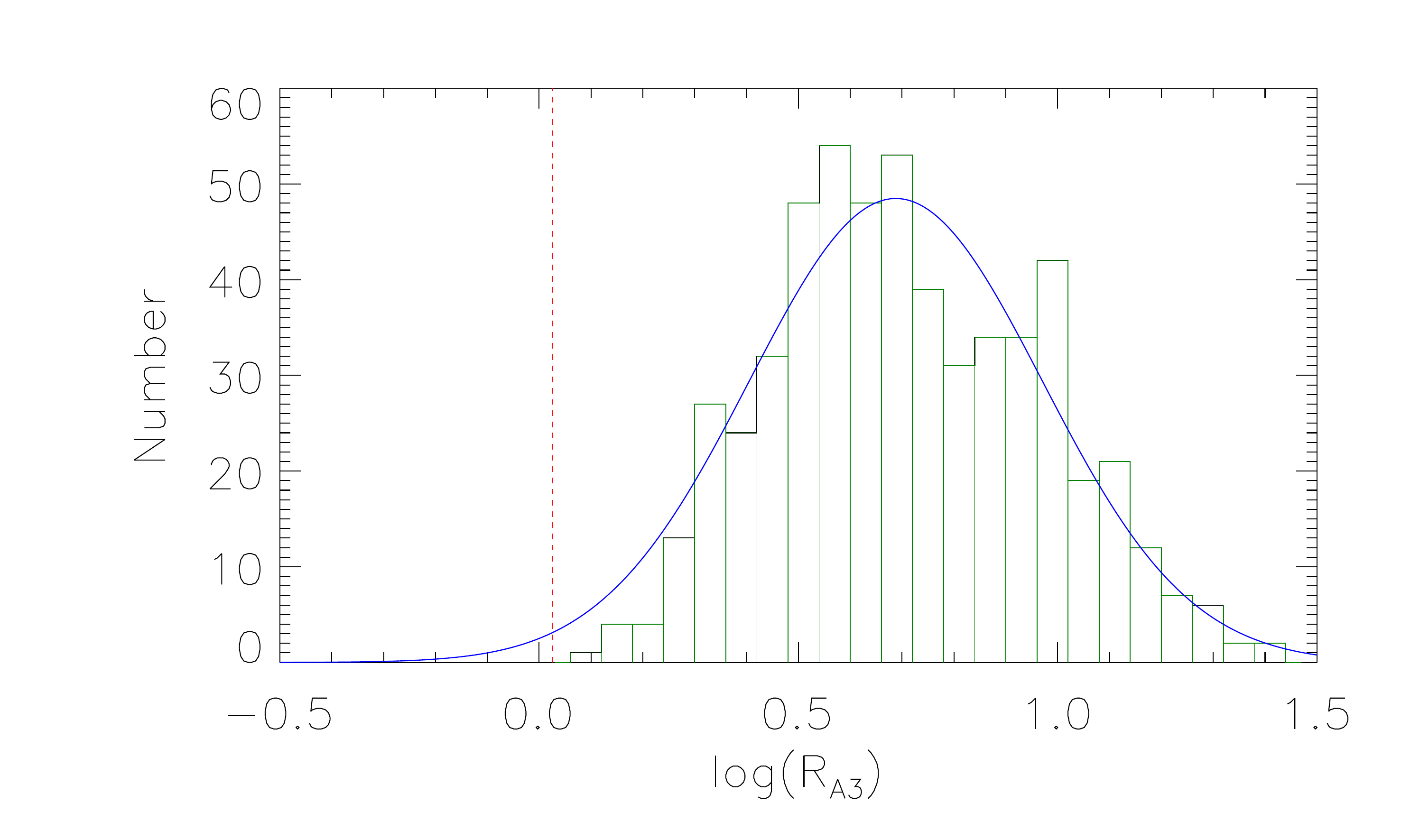}
\caption{Distribution of line width ratio $\log(R_{A3})$ of broad H$\alpha$ to extended component in 
[O~{\sc iii}]$\lambda5007$\AA~  of the 535 QSOs reported in \citet{zh21a}. Vertical dashed red line shows 
$R_{A3}=1.06$ for \obj. Solid blue line shows the Gaussian description to the distribution of $\log(R_{A3})$.}
\label{rat}
\end{figure}

\begin{table*}
\caption{Shifted velocities $SV$ listed in this Letter}
\begin{tabular}{|p{3cm}|p{14cm}|}
\hline\hline
value    &  description   \\
\hline
	$SV_{SSP}=500\pm20$km/s  &  SSP method determined velocity shift for the Stellar Templates to describe absorption 
	features in the SDSS spectrum with respect to $z\sim0.283$ \\
\hline
$SV_{B,B}=690\pm240$km/s  & Shifted velocity for the broad component in H$\alpha$, calculated by wavelength difference between
	the measured central wavelength of broad H$\alpha$ in rest frame ($z\sim0.283$) and the  
	theoretical vacuum wavelength 6564.61\AA \\
\hline
	$SV_{B,O3}=980\pm50$km/s  & Shifted velocity for the broad component in [O~{\sc iii}], wavelength difference 
	between the measured central wavelength of the broad component in [O~{\sc iii}]$\lambda5007$\AA~ in rest frame 
	($z\sim0.283$) and the theoretical vacuum wavelength 5008.24\AA\\
\hline
	$SV_{N2}=370\pm75$km/s  & Shifted velocity for the extended component in [N~{\sc ii}], relative to the core 
	component in [N~{\sc ii}]$\lambda6583$\AA \\
\hline\hline
\end{tabular}
\end{table*}


	Furthermore, in Fig.~\ref{rat}, we show the distribution of the line width ratio $R_{A3}$ between broad H$\alpha$ 
(H$\alpha_B$) and broad extended component (O3E) in [O~{\sc iii}] line for the 535 normal QSOs \citep{zh21a} with a mean value of 
$\log(R_{A3})\sim0.71\pm0.26$ (the standard deviation accepted as the uncertainty 0.26). The dashed vertical red line indicates the 
value for SDSS J1619 ($\sigma_{H\alpha_B}/\sigma_{O3E}=900/850\sim1.06$). There is no QSO with $R_{A3}$ smaller than 
the value for SDSS J1619, which implies that the probability of detecting QSOs with $R_{A3}<1.06$ is lower than $1.87\times10^{-3}$ (1/535) and suggests that SDSS J1619 is a unique source.

	Finally, four velocity shifts measured for SDSS 1619 based on the SDSS redshift ($z=0.283$) are listed in Table 2 with 
clear descriptions. And the velocity shifts $SV$ of the broad components in H$\alpha$ and [O~{\sc iii}] are mainly considered in 
the next section. If we adopt the stellar absorption feature as a reliable redshift of SDSS J1619 ($z=0.285$), the measured velocity 
shifts in Table 2 increase ($SV_{B,B}\sim1080~{\rm km/s}$ and $SV_{B,O3}\sim1370{\rm km/s}$) but the line widths do not change.

\section{Discussion}

\begin{figure}
\centering\includegraphics[width = 8cm,height=5cm]{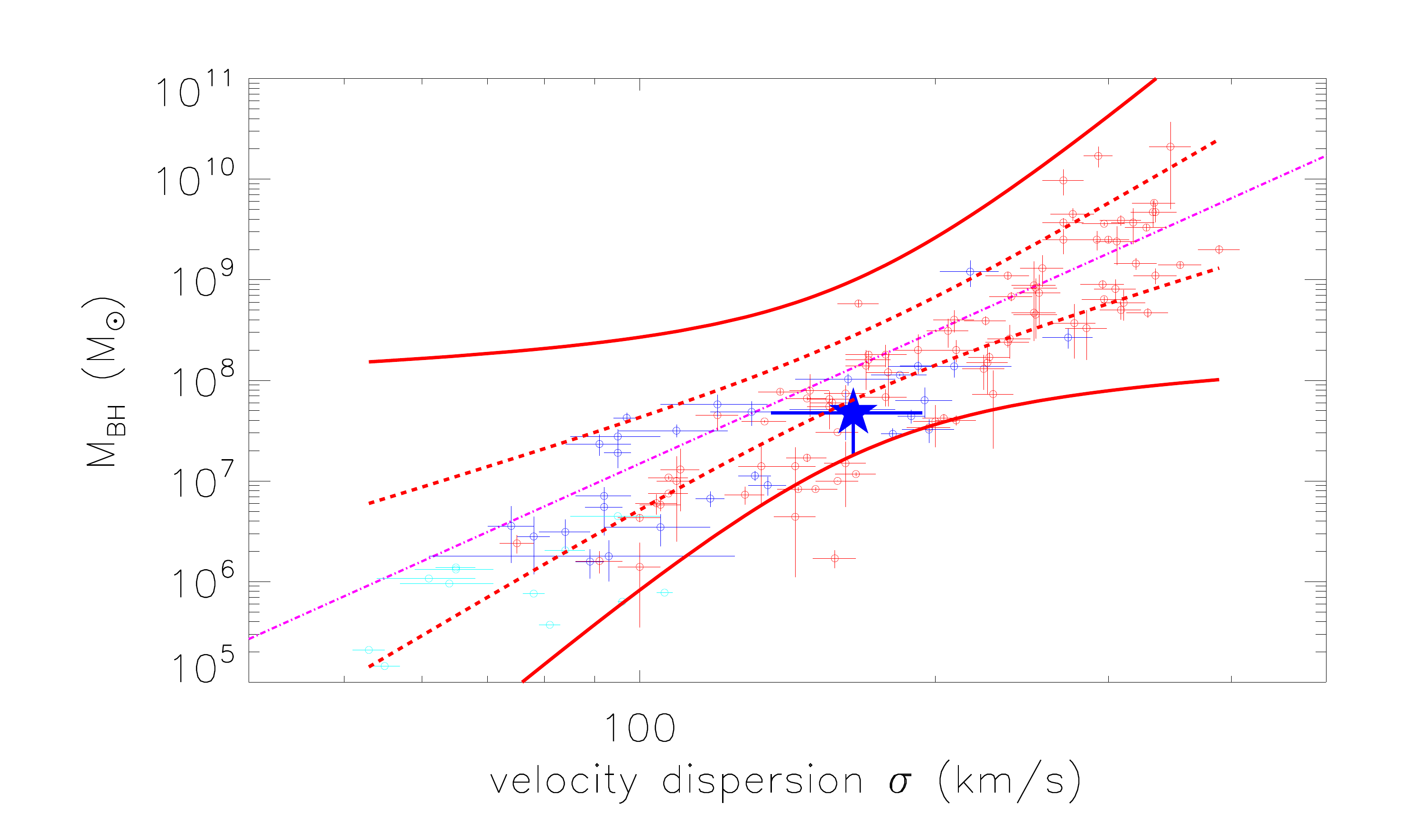}
\caption{On dependence of BH mass on stellar velocity dispersion. Solid five-point-star with error bars in blue show the $M_{BHr}$ 
and the velocity dispersion and corresponding uncertainties of SDSS J1619. Open circles in red, in blue and in cyan 
show the values for the 89 quiescent galaxies from \citet{sg15}, the 29 RM AGNs from \citet{wy15} and the 12 TDEs from \citet{zl21}, 
respectively. Dot-dashed pink line, dashed red and solid red lines show the relation reported in \citet{kh13} and 
the corresponding 3$\sigma$ and 5$\sigma$ confidence bands.}
\label{msig}
\end{figure}

	Based on the best fitting results, a blue-shifted broad component can be found in H$\alpha$. However, it is necessary 
to determine where do the determined blue-shifted broad component in H$\alpha$ come from, from emission regions in NLRs or 
in BLRs.

	Adopting the intrinsic flux ratio 3.1 of broad H$\alpha$ to broad H$\beta$, the observed flux ratio 
$12.35_{-5.38}^{+11.12}$ of broad H$\alpha$ to broad H$\beta$ in \obj~ indicates severe obscurations on the broad 
Balmer emission lines have $E(B-V)\sim1.22_{-0.49}^{+0.55}$, leading the intrinsic broad blue-shifted component in H$\alpha$ to 
have line flux about $4405_{-3025}^{+1054}\times10^{-17}{\rm erg/s/cm^2}$ (corresponding intrinsic line luminosity about 
$11.11_{-7.61}^{+2.66}\times10^{42}{\rm erg/s}$). If accepted the blue-shifted broad component in H$\alpha$ tightly related 
to BLRs bound to the expected rSMBH in \obj, based on the virialization assumption \citep{ve02, pe04, sh11, mt22}, 
combining with the second moment 19.76$\pm$3.35\AA~ of the blue-shifted broad H$\alpha$ and the improved empirical 
dependence \citep{bd13} of BLRs size on the intrinsic broad line luminosity \citep{gh05}, virial mass of the recoiling BH in \obj~ 
can be estimated by 
\begin{equation}
\begin{split}
M_{BHr}&=15.6\times10^6(\frac{L_{H\alpha}}{\rm 10^{42}erg/s})^{0.55}
	(\frac{\sigma_{H\alpha}}{\rm 1000km/s})^{2.06}{\rm M_\odot} \\
	&=4.75_{-3.04}^{+2.64}\times10^7{\rm M_\odot}
\end{split}
\end{equation}
with uncertainties determined by the uncertainties of line width and intrinsic line luminosity of the blue-shifted broad 
component in H$\alpha$.

	If the blue-shifted broad component in H$\alpha$ was not related to normal BLRs, the estimated $M_{BHr}$ should be quite 
different from the BH mass expected from the \msig~ relation in \obj. The \msig~ relation discussed in \citet{fm00, ge00, kh13, 
bb17, bt21} can be conveniently applied to estimate central BH mass in both quiescent galaxies and active galaxies. Moreover, as 
discussed in \citet{ds05, jn09} for BBH systems at sub-pc scales, central total BH mass could be also estimated by the \msig~ 
relation. Therefore, it is interesting to check properties of the virial BH mass of \obj~ in the plane of BH mass versus stellar 
velocity dispersion in Fig.~\ref{msig}. Here, in order to show clearer results in Fig.~\ref{msig}, not only the \msig~ relation 
reported in \citet{kh13} is shown, but also the 89 quiescent galaxies from \citet{sg15} and the 29 reverberation mapped (RM) AGN 
from \citet{wy15} and the 12 tidal disruption events (TDEs) from \citet{zl21} are also shown in the figure. Interestingly, in 
Fig.~\ref{msig}, $M_{BHr}$ of \obj~ is consistent with the mass expected from the \msig~ relation, based on the measured 
stellar velocity dispersion 165$\pm$29km/s in \obj. Therefore, the blue-shifted broad component in H$\alpha$ is from emission 
regions related to BLRs, and a rSMBH can be applied to explain the blue-shifted broad component in H$\alpha$.

	Moreover, as discussed in \citet{ms06, gm08, km08b}, materials can be bound to a rSMBH within a region with radius 
$r_k$ given by
\begin{equation}
r_k~\sim~512\frac{M_{BH}}{\rm 10^8M_\odot}(\frac{V_{k}}{\rm 10^3km/s})^{-2} {\rm light-days} 
\end{equation}
with $M_{BH}$ and $V_k$ as the BH mass and the kick velocity of a rSMBH. In \obj~, based on the continuum luminosity at 5100\AA~ 
about $18.8_{-14.7}^{+85.2}\times10^{44}{\rm erg/s}$ after considering the severe obscuration with 
$E(B-V)\sim1.22_{-0.49}^{+0.55}$, 
the expected BLRs size is about $184_{-108}^{+261}$light-days. Considering the $M_{BHr}\sim{\rm 4.75\times10^7M_\odot}$ in \obj, 
$V_{k}\sim900$km/s (the observed velocity shift) can lead to $r_k\sim500$light-days, larger enough than the estimated 
BLRs size. In other words, the materials bound to the expected rSMBH in \obj~ are enough to be applied to estimate the  
virial BH mass. Meanwhile, if accepted the new redshift 0.285 through the stellar absorption features, $V_{k}$ should 
be corrected to $V_k\sim1300$km/s, leading to $r_k\sim250$light-days, still larger than the estimated BLRs size.

	Therefore, considering the amplitude of parsecs to kilo-pcs of a rSMBH, when the rSMBH is wandering through the NLRs 
in \obj, the blue-shifted broad components in [O~{\sc iii}] doublet naturally have the same velocity shifts  
and the same line widths as those of the blue-shifted broad H$\alpha$ from the BLRs bound to the rSMBH. Before end of the section, 
besides the preferred rSMBH scenario in \obj, the other three probable models are simply discussed as follows. 

	First and foremost, there are double-peaked features in [O~{\sc iii}]$\lambda5007$\AA~ doublet widely accepted as signs 
of dual core systems at kilo-pc scales \citep{zw04, xk09, fz11, wl19}. Based on the double-peaked features in 
[O~{\sc iii}]$\lambda5007$\AA, the peak separation is about 491$\pm$31km/s, leading broad emission lines from central two cores 
to have the same peak separation 491$\pm$31km/s. However, the velocity shift about 900km/s of the blue-shifted 
broad H$\alpha$ in \obj~ is about two times ($z\sim0.283$ or $z\sim0.285$ accepted) higher than the 491$\pm$31km/s, indicating the 
blue-shifted broad H$\alpha$ is not related to a dual core system expected by the double-peaked [O~{\sc iii}] in \obj. Meanwhile, 
if a BBH system was assumed in \obj~ without considerations of the double-peaked features in [O~{\sc iii}]$\lambda5007$\AA, 
a blue-shifted broad H$\alpha$ could be expected in \obj. However, it is hard to explain the blue-shifted broad 
components in [O~{\sc iii}]$\lambda5007$\AA~ and in H$\alpha$ have the same line widths, as shown above that the probability is 
smaller than $1.87\times10^{-3}$ to detect an AGN with $R_{A3}\le0.94$. Therefore, a BBH system is disfavoured in \obj. 

	Besides, a disk emitter \citep{ch89, el95, st03} can be applied to explain the blue-shifted broad H$\alpha$, however, 
it is also hard to explain the blue-shifted broad H$\alpha$ has the same line width as that of the blue-shifted broad component 
in [O~{\sc iii}]$\lambda5007$\AA~ in \obj, due to lower probability $1.87\times10^{-3}$ to find an AGN with $R_{A3}\le0.94$. 
Therefore, the BBH model is not favoured in \obj.

	Last but not the least, an outflowing model could be applied to explain the similar velocity shifts and 
the similar line widths of the broad components in H$\alpha$ and [O~{\sc iii}] doublet, if accepted the broad components 
in H$\alpha$ and in [O~{\sc iii}] doublet were from NLRs. If the broad components were not from BLRs but from NLRs, it is hard to 
expect the consistency between the virial $M_{BHr}$ and the mass from the \msig~ relation, while considering the probability 
lower around $1.87\times10^{-3}$ to find an AGN with $R_{A3}\le0.94$. Therefore, the outflowing model is disfavoured in \obj.

\section{Conclusions}

	After considering the advantages of studying rSMBH in Type-1.9 AGN, similar line widths and velocity shifts 
of the blue-shifted broad components in H$\alpha$ and [O~{\sc iii}] doublet are reported in the Type-1.9 AGN SDSS J1619. 
Based on the consistency between the central virial BH mass in SDSS J1619 and the BH mass expected from the \msig~ relation, the 
blue-shifted broad H$\alpha$ can be accepted to come from BLRs bound to the expected rSMBH in SDSS J1619. Meanwhile, the 
expected rSMBH wandering through NLRs can be naturally applied to explain the similar velocity shifts and the similar line 
widths of the blue-shifted broad component in H$\alpha$ and in [O~{\sc iii}] in SDSS J1619. Therefore, the rSMBH scenario is 
preferred in \obj.

\section*{Acknowledgements}
ZXG gratefully acknowledges the anonymous referee for reading our manuscript carefully and patiently, and giving us constructive 
comments and suggestions to greatly improve our paper. ZXG gratefully thanks the grant support from research funding by GuangXi 
University and the grant support from NSFC-12173020 and NSFC-12373014. The Letter has made use of the data from the SDSS projects 
with web site \url{http://www.sdss3.org/}. The Letter has made use of the MPFIT package 
(\url{http://cow.physics.wisc.edu/~craigm/idl/idl.html}) written by Craig B. Markwardt. 

\section*{Data Availability}
The data underlying this article will be shared on reasonable request to the corresponding author
(\href{mailto:xgzhang@gxu.edu.cn}{xgzhang@gxu.edu.cn}).

\label{lastpage}
\end{document}